\documentclass[english,reprint]{revtex4}
\usepackage[T1]{fontenc}
\usepackage[latin9]{inputenc}
\usepackage{textcomp}
\usepackage{amstext}
\usepackage{amssymb}
\usepackage{graphicx}
\usepackage{esint}

\makeatletter

\DeclareRobustCommand{\greektext}{%
  \fontencoding{LGR}\selectfont\def\encodingdefault{LGR}}
\DeclareRobustCommand{\textgreek}[1]{\leavevmode{\greektext #1}}
\DeclareFontEncoding{LGR}{}{}
\DeclareTextSymbol{\~}{LGR}{126}
\newcommand{\lyxmathsym}[1]{\ifmmode\begingroup\def\b@ld{bold}
  \text{\ifx\math@version\b@ld\bfseries\fi#1}\endgroup\else#1\fi}

\@ifundefined{textcolor}{}
{%
 \definecolor{BLACK}{gray}{0}
 \definecolor{WHITE}{gray}{1}
 \definecolor{RED}{rgb}{1,0,0}
 \definecolor{GREEN}{rgb}{0,1,0}
 \definecolor{BLUE}{rgb}{0,0,1}
 \definecolor{CYAN}{cmyk}{1,0,0,0}
 \definecolor{MAGENTA}{cmyk}{0,1,0,0}
 \definecolor{YELLOW}{cmyk}{0,0,1,0}
 }

\usepackage{babel}

\usepackage{babel}

\usepackage{babel}

\usepackage{babel}

\usepackage{babel}

\makeatother

\usepackage{babel}
\begin{document}

\title{Ordered arrays of magnetic nanowires investigated by polarized small-angle
neutron scattering}

\author{Thomas Maurer}

\affiliation{Laboratoire de Nanotechnologie et d\textquoteright{}Instrumentation
Optique, ICD CNRS UMR STMR 6279, Université de Technologie de Troyes,
12 rue Marie Curie, CS 42060, 10004 Troyes Cedex, France}

\author{Sébastien Gautrot, Frédéric Ott, Grégory Chaboussant }

\email{gregory.chaboussant@cea.fr}

\selectlanguage{english}%

\affiliation{Laboratoire L\textasciiacute{}eon Brillouin, UMR12 CEA-CNRS, 91191
Gif-sur-Yvette, France}

\author{Fatih Zighem}

\affiliation{LSPM, CNRS-Université Paris XIII, Sorbonne Paris Cité, 93430 Villetaneuse,
France}

\author{Laurent Cagnon, Olivier Fruchart}

\affiliation{Institut N\textasciiacute{}eel, CNRS et Université Joseph Fourier,
BP166, F-38042 Cedex 9 Grenoble, France}
\begin{abstract}
Polarized small-angle neutron scattering (PSANS) experimental results
obtained on arrays of ferromagnetic Co nanowires ($\phi\approx13$
nm) embedded in self-organized alumina (Al$_{2}$O$_{3}$) porous
matrices are reported. The triangular array of aligned nanowires is
investigated as a function of the external magnetic field with a view
to determine experimentally the real space magnetization $\vec{M}(\vec{r})$
distribution inside the material during the magnetic hysteresis cycle.
The observation of field-dependentSANSintensities allows us to characterize
the influence of magnetostatic fields. The PSANS experimental data
are compared to magnetostatic simulations. These results evidence
that PSANS is a technique able to address real-space magnetization
distributions in nanostructured magnetic systems. We show that beyond
structural information (shape of the objects, two-dimensional organization)
already accessible with nonpolarized SANS, using polarized neutrons
as the incident beam provides information on the magnetic form factor
and stray fields \textgreek{m}0Hd distribution in between nanowires.
\end{abstract}


\maketitle

\section{Introduction }

The structural, magnetic, and optical properties of nanoobjects organized
in periodic arrays have been intensively studied in recent years,
as part of the growing interest in functionalized magnetic nanostructures.
Several converging lines of effort have greatly improved our knowledge
of magnetic nano-objects over the last years. It started with the
development of a wide range of systems, from dots to wires, with well
controlled structural and magnetic features, and foreseen applications
in medicine and magnetoelectronics {[}1\textendash{}3{]}.

In this respect, \textquotedblleft{}elongated\textquotedblright{}
magnetic nano-objects in the form of nanowires (or nanorods) with
very high aspect ratio (length/radius) have emerged as some of the
most promising materials due to several factors {[}4{]}. First, synthesis
improvements based on self-organization principles have made it possible
to produce arrays of very high quality with narrow size distribution
{[}5{]} and two-dimensional (2D) organization inmatrices {[}6,7{]}.
Second, themagnetic properties of ferromagnetic (FM) nanowires are
essentially governed by their shape anisotropy, leading to large magnetic
coercivity {[}8,9{]}, and hence potential for electronic devices or
high-density storage, and even high temperature permanent magnets
{[}10{]}. The understanding of these nanosized systems calls for advanced
characterization techniques with high sensitivity and spatial resolution
{[}11{]} such as a-SNOM (\textquotedblleft{}apertureless\textquotedblright{}
scanning near-field optical microscopy) {[}12\textendash{}14{]}, spin-polarized
STM {[}15{]}, electronic holography {[}16{]}, XMCD-PEEM (x-ray magnetic
circular dichroism with photoemission electromicroscopy) {[}17{]},
SPLEEM (spin-polarized low-energy electron microscopy) {[}18{]}, etc.
Most of these techniques are realspace and local, so that complementary
approaches should be developed to address the properties of large
assemblies of nano-objects either deposited on surfaces or buried
in layers. Raman spectroscopy, Brillouin light scattering, x-ray scattering,
and neutron scattering (diffraction for structures, inelastic for
excitations, and SANS for large-scale objects) are techniques of choice
in this regard. The latter technique benefits greatly from a wide
available $q$ range. Finally, large theoretical and numerical efforts
{[}19\textendash{}21{]} to address fundamental issues related to the
transition from an atomic description to a \textquotedblleft{}nanoscale\textquotedblright{}
description, from discrete approaches to continuous models, have helped
us understand the collective behavior of nanoscale objects, in particular
magnetic nanowires {[}22\textendash{}24{]}.

In this article,we report small-angle neutron scattering with polarized
neutrons (PSANS) investigations of ferromagnetic Co nanowires embedded
in an alumina matrix. This technique is well suited to probe both
the size and shape of nano-objects through the characterization of
the form factors (magnetic and nuclear), and their spatial organization
through the structure factor. Among all the synthesis strategies,
nanowires electrochemically grown inside porous alumina matrices are
ideal candidates to test PSANS as they produce triangular arrays with
long-range coherence. Only recently, there were reported PSANS studies
of magnetic nanowires {[}25\textendash{}27{]}. In particular, Napolskii
et al. {[}27{]} suggested that the magnetic contribution to SANS could
only be explained by taking into account magnetostatic fields between
the nanowires.We show in this paper that using polarized neutrons
as the incident beam provides information on the magnetic form factor
and stray fields $\mu_{0}H_{d}$ distribution.

\section{Cobalt magnetic nanowires: synthesis and bulk magnetic properties}

\begin{figure}
\includegraphics[bb=0bp 150bp 842bp 595bp,clip,width=8.5cm]{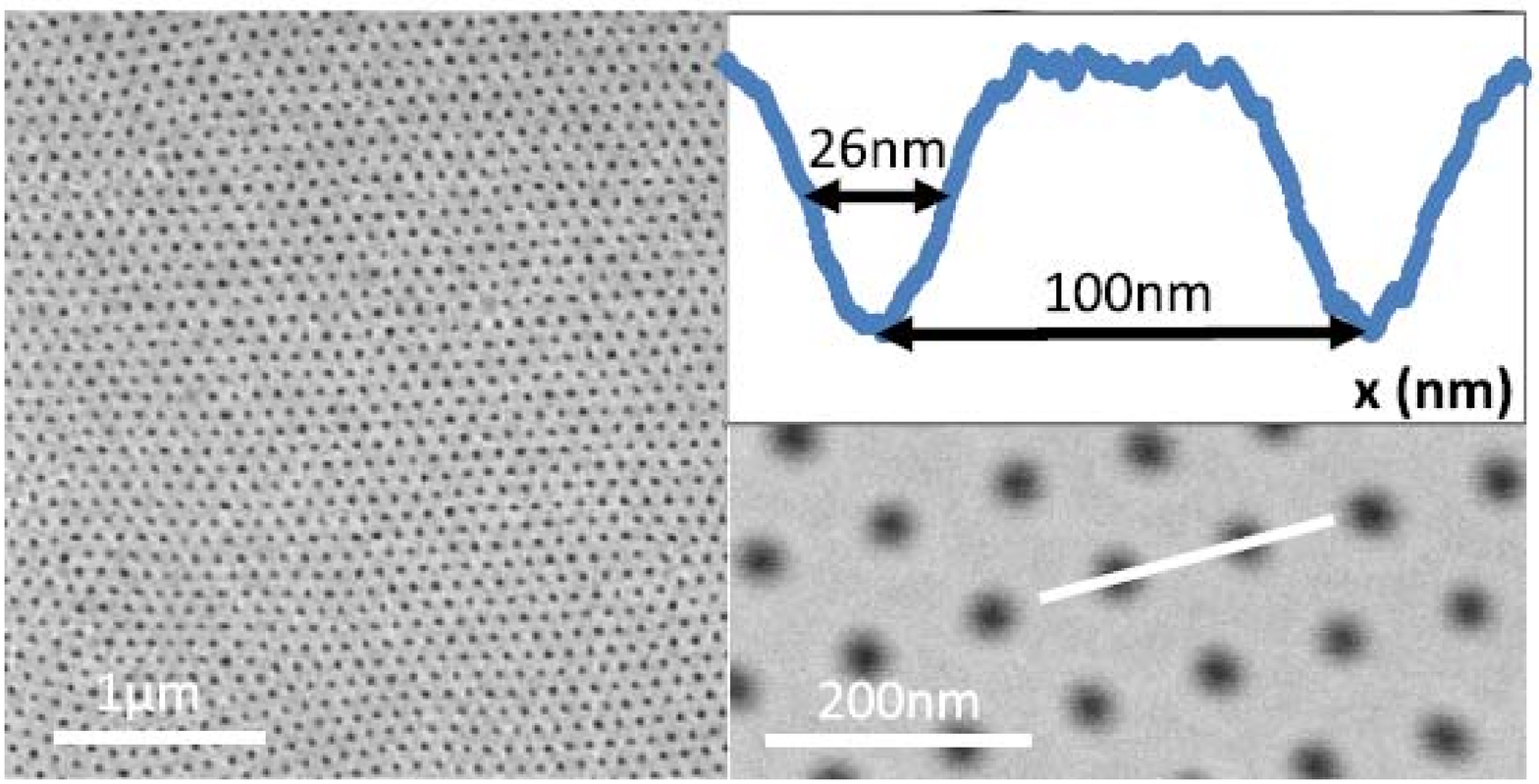}
\caption{SEM images of the porous alumina membrane used to produce the Co nanowires,
and extraction by contrast image treatment of the depth profile of
the pores as seen by SEM. The apparent average diameter is estimated
to be around 26 nm, but the \textquotedblleft{}surface\textquotedblright{}
diameter is closer to 50 nm, reflecting both a potential conical shape
of the pores and also a bias due to the techniques employed to extract
the size (image contrast).}

\label{Fig1:images} 
\end{figure}

The porous alumina membranes are first formed using a double anodization
process. The synthesis starts from a bulk Al plate which is electrochemically
oxidized to form alumina, Al$_{2}$O$_{3}$. During the process, the
alumina layer forms self-organized, spatially ordered nanopores whose
diameters and interpore distances vary from a few tens of nanometers
up to hundreds of nanometers. The different steps involved in the
synthesis of porous alumina membranes are extensively documented in
Refs. {[}28\textendash{}34{]}. Once the alumina porous layer is formed
and after Au deposition (150 nm in thickness) on one surface to act
as an electrode, one can fill by electrodeposition the pores with
various 3$d$ metals (Fe, Co, Ni) and alloys (CoPt, FePt, FeNi, CoNi)
to obtain an array of nano-objects with diameters in the range 10-100
nm and lengths up to several tens of $\mu$m {[}35,36{]}. The growth
process is controlled either by the applied current (galvanostatic
deposition) or the applied potential (potentiostatic deposition),
and influenced by electrolyte pH {[}6,37{]}.

In particular, it is possible to synthesize hcp Co nanowires with
the $\vec{c}$ axis either parallel or perpendicular to the long axis
of the nanowires (for hcp, the preferential growth has the c axis
perpendicular to the nanowire axis and, in such a case, there is a
competition between the magneto-crystalline anisotropy field and the
shape anisotropy field leading to a global coercivity decrease). From
scanning electron microscopy (SEM) images, the pores are characterized
by their average apparent diameter $\phi_{P}=26$ nm and interpore
distance $d_{P}=105$ nm (see Fig. 1). We argue below that SANS studies
may give slightly different results due to the fact that SANS averages
the signal over the depth of the wires whereas SEM images give a blurred
surface view of the top of the nanowires.

\begin{figure}
\includegraphics[bb=0bp 60bp 842bp 595bp,clip,width=8.5cm]{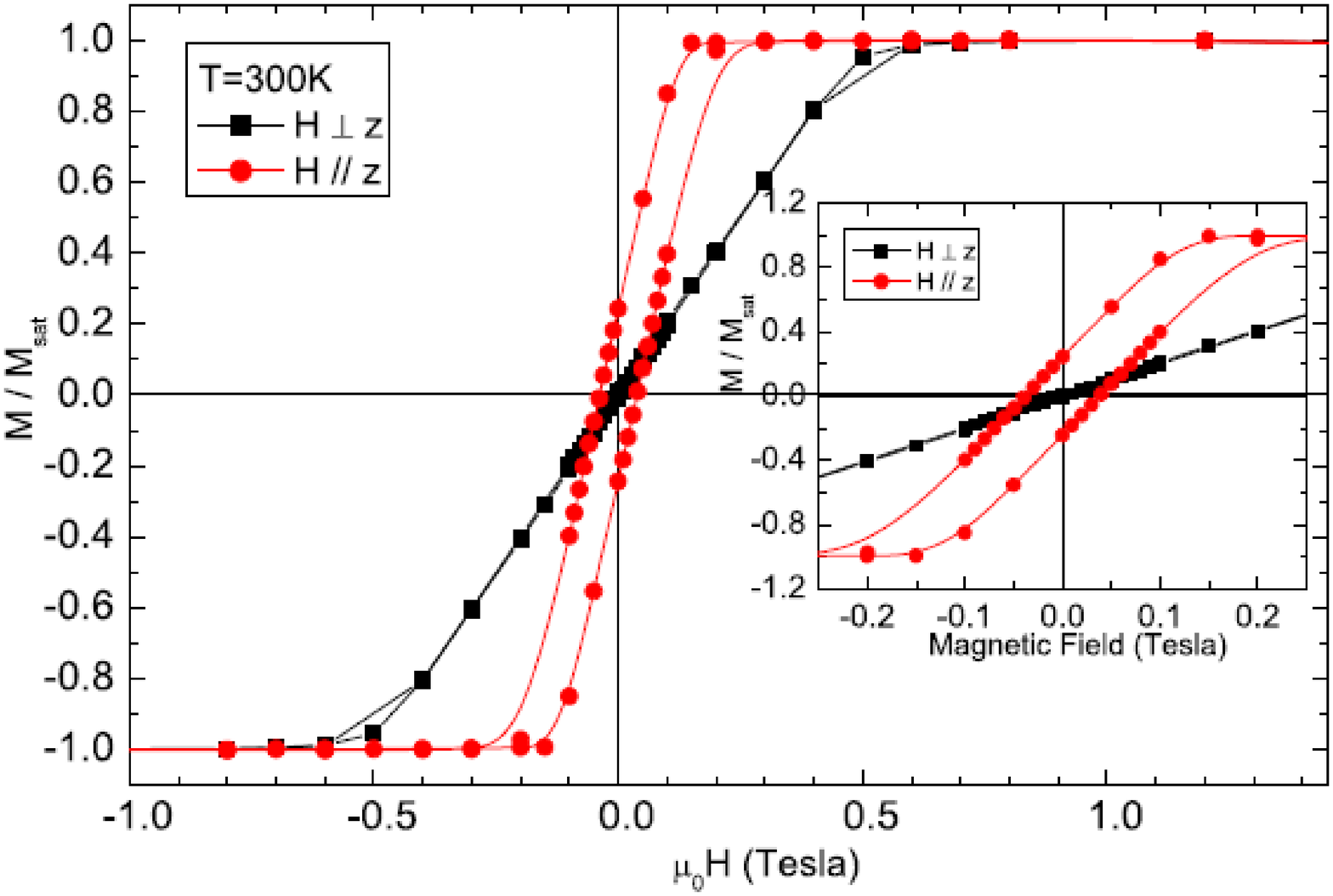}
\caption{Room temperature VSM (vibrating sample magnetometer) measurements
of an array of amorphous Co nanowires (thiocyanate SCN$^{-}$) in
their alumina matrix measured along the nanowire axis ($\vec{H}\parallel\vec{z}$)
and in the plane of the membrane ($\vec{H}\perp\vec{z}$): $\mu_{0}H_{C}^{\parallel}=38$
mT, $M_{R}^{\parallel}=0.25M_{S}$ and $\mu_{0}H_{C}^{\perp}=3.7$
mT, $M_{R}^{\perp}=0.008M_{S}$.}
\end{figure}

In our case, Co nanowires have been electrodeposited from a solution
based on boric acid in which cobalt sulfate and potassium thiocyanate
salts have been added. The length of the produced Co nanowires is
about 15 $\mu$m and they exhibit an amorphous and/or nanocrystallized
structure due to the incorporation of thiocyanate ions (known to strongly
interact with noble and transition metals {[}38{]}) during the growth.
Therefore, magnetocrystalline anisotropy is expected to be absent,
and only the wire and sample global shape anisotropy will contribute
to the hysteresis loop {[}39{]}. Transmission electron microscopy
(TEM) images show that the pores are well filled and the Co part of
the nanowires is homogeneous, which tends to prove that the scattering
length density (SLD) can be considered constant across the section
of the wires. Magnetic hysteresis cycles for magnetic fields applied
parallel ($\vec{z}$ direction) or perpendicular to the wires' axes
are presented in Fig. 2. For $\vec{H}\perp\vec{z}$, the hysteresis
cycle is closed ($\mu_{0}H_{C}^{\perp}=3.7$ mT and $M_{R}^{\perp}=0.008M_{S}$).
For $\vec{H}\parallel\vec{z}$, the remanence is only $M_{R}^{\parallel}=0.25M_{S}$
and the coercive field is $\mu_{0}H_{C}^{\parallel}=38$ mT. Under
the assumption that each wire is at any moment mostly uniformly magnetized,
and switches abruptly with a finite coercivity, the reduced remanence
$M_{R}^{\parallel}$ and slanted shape of the loop results from the
interwire magnetostatic interactions, whose strength is to first order
measured by the slanting {[}40\textendash{}42{]}. Domain formation
at lowfields is possible and cannot be a priori discarded; however,
the actual model to describe the SANS data shown hereafter will still
stand even in this case, as only the amplitude of the magnetic form
factor may change and not the structure factor nor the nuclear form
factor.

\section{PSANS intensity: form and structure factors}

\begin{figure}
\includegraphics[bb=30bp 210bp 842bp 595bp,clip,width=8.5cm]{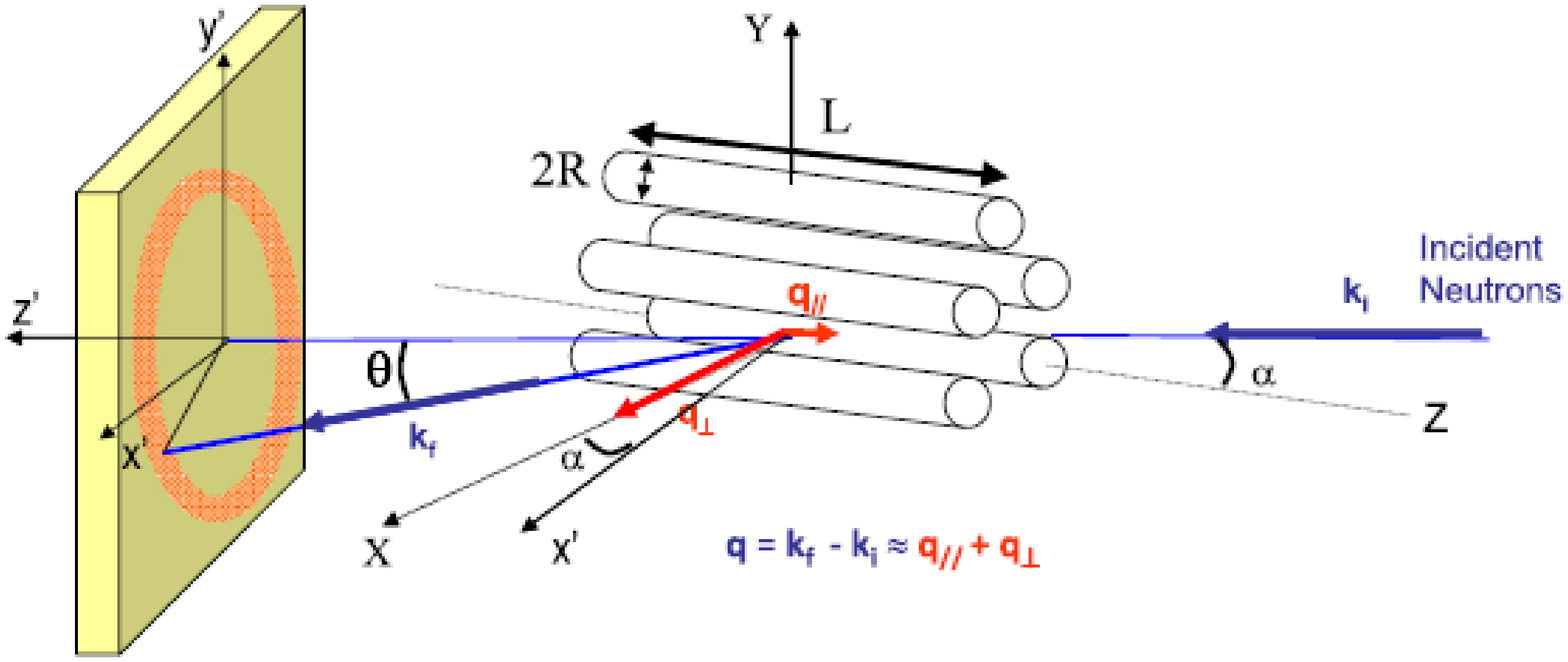}
\caption{Schematics of the SANS experiments on oriented nanowires (length $L$
and radius $R$). $\alpha$ is defined as the angle between the cylinder
axis ($Z$ axis) and the incident beam direction $\vec{k}_{i}$ ($k_{i}=2\lyxmathsym{\textgreek{p}}/\lambda$).
The scattering vector $\vec{q}=\vec{k}_{f}-\vec{k}_{i}\approx k_{i}\sin\theta$
can be expressed in the small-angle approximation {[}$\cos\theta\approx1${]}
as $\vec{q}=\vec{q}_{\parallel}-\vec{q}_{\perp}$ with $\vec{q}_{\parallel}=q\sin\alpha\hat{Z}$
(component along $Z$) and $\vec{q}_{\parallel}=q\cos\alpha\hat{X}$
(component along $X$).}
\end{figure}

\subsection{General expressions}

The PSANS intensity $I^{\pm}(q)$ ($\pm$ stands for up or down incoming
polarization with respect to the applied magnetic field) for a perfectly
polarized beam may be written with good approximation as: 
\begin{equation}
I^{\pm}(q)=\left|F_{N}(q)\pm F_{M}(q)\right|^{2}S(q)
\end{equation}

where $F_{N}(q)$, $F_{M}(q)$ and $S(q)$ are respectively the nuclear
and magnetic form factors, and the structure factor, of the scattering
objects. The scattering vector modulus $q$ is defined as $q=k_{i}\sin\theta$,
where $k_{i}=2\lyxmathsym{\textgreek{p}}/\lambda$ and $\theta$ denotes
the scattering angle (see Fig. 3). The structure factor $S(q)$ gives
access to the spatial correlations between the objects. The nuclear
form factor $F_{N}(q)$ depends on the nuclear scattering length density
(SLD) contrast $\rho$ between the different chemical elements present
in the sample (wires plus matrix) and on the geometrical form factor
$F_{geo}(q)$ which is governed by the shape of the objects. The magnetic
form factor $F_{M}(q)$ has an expression related to the local magnetization
$\vec{M}(\vec{r})$ which will be detailed below. We define $\Sigma I(q)$
and $\Delta I(q)$ as:
\begin{equation}
\Sigma I(q)=I^{+}(q)+I^{-}(q)=2\left[F_{N}^{2}(q)+F_{M}^{2}(q)\right]S(q)
\end{equation}

and 
\begin{equation}
\Delta I(q)=I^{+}(q)-I^{-}(q)=4F_{N}(q)F_{M}(q)S(q)
\end{equation}

In a real situation, the neutron polarization ratio $P$ is smaller
than 1 (usually in the range $0.9-0.95$ depending on the setup) and
thus we must consider mixing channels between up and down beams so
that the expected $I_{r}^{+}(q)$ and $I_{r}^{-}(q)$ become:
\begin{equation}
I_{r}^{+}(q)=\frac{1}{2}\left(1+P\right)I^{+}(q)+\frac{1}{2}\left(1-P\right)I^{-}(q)
\end{equation}
 and 
\begin{equation}
I_{r}^{-}(q)=\frac{1}{2}\left(1-P\right)I^{+}(q)+\frac{1}{2}\left(1\text{+}P\right)I^{-}(q)
\end{equation}

\subsection{Nuclear form factors $F_{N}(q)$}

In the case of particles of volume VP with a SLD $\rho_{P}$ dispersed
in a medium with SLD \textgreek{r}med, the nuclear form factor $F_{N}(q)$
of the particles is defined as:

\begin{equation}
F_{N}(\vec{q})=\left(\rho_{P}-\rho_{med}\right)\intop_{V_{P}}e^{-i\vec{q}\cdot\vec{r}}d\vec{r}=(\Delta\rho)F_{geo}(\vec{q})
\end{equation}

where $F_{geo}(\vec{q})$ is the geometrical form factor for one single
nanowire, depending only upon the shape and dimensions of the particle,
and $\Delta\rho=\rho_{P}-\rho_{med}$.

\subsubsection{Uniform cylinder}

The geometrical form factor $F_{geo}(\vec{q})$ of one nanowire is,
to a very good approximation, equal to that of a filled cylinder of
radius $R$ and length $L$. Assuming the nanowire axis, defined as
the $Z$ axis, makes an angle $\alpha$ with respect to the incident
beam direction $\vec{k}_{i}$, then the geometrical form factor of
a cylinder is expressed as {[}43\textendash{}45{]}:
\begin{equation}
F_{geo,cyl}(\overrightarrow{q})=V_{P}\frac{2J_{1}(q_{\perp}R)}{q_{\perp}R}.\frac{\sin(\frac{1}{2}q_{\shortparallel}L)}{\frac{1}{2}q_{\shortparallel}L}
\end{equation}

where $J_{1}(x)$ is the first-order Bessel function. The longitudinal
and transverse components of $\vec{q}$ ($=\vec{q}_{\parallel}+\vec{q}_{\perp}$)
are $\vec{q}_{\parallel}=q\sin\alpha\hat{Z}$ and $\vec{q}_{\perp}=q\cos\alpha\hat{X}$
(see Fig. 3). In the case of perfect alignment between $\vec{k}_{i}$
and the $Z$ axis ($\alpha=0,q_{\parallel}=0$), $F_{geo,cyl}(\overrightarrow{q})$
is simply:
\begin{equation}
F_{geo,cyl}(\overrightarrow{q},\alpha=0)=V_{P}\frac{2J_{1}(q_{\perp}R)}{q_{\perp}R}
\end{equation}
This is similar to the scattering of a flat disk, as the length $L$
of the nanowires has no influence. The alignment process is an important
aspect of SANS experiments on elongated ordered objects such as nanowires.
For imperfect alignment or radius variation (roughness, interwire
distributions), the expression must take into account the length $L$,
the dispersion in $\alpha$, and $R$ values as shown by Pépy et al.
(SAXS) {[}43{]} and Marchal et al. (SANS) {[}44{]}. Noticeably, the
length $L$ will play a role in the stray field spatial distribution
(and thus on the PSANS intensity) in a regime where the aspect ratio
$L/D=1\lyxmathsym{\textminus}10$, but this is not the case here ($L/D\geq100$),
where we have extremely homogeneous stray field distribution (see
Fig. 2 in Ref. {[}46{]}).

\subsubsection{Core-shell cylinder}

The form factor for a \textquotedblleft{}core-shell\textquotedblright{}
cylinder (defined by a cylinder of core radius $R$ and core length
$L$ with shell thickness $t$ and total length $L+2t$) is given
by {[}47{]}:
\begin{equation}
F_{N}(\overrightarrow{q})=2\left(\rho_{core}-\rho_{shell}\right)V_{core}J_{0}\left(q\frac{L}{2}\sin\alpha\right)\frac{J_{1}(u)}{u}+2\left(\rho_{shell}-\rho_{med}\right)V_{shell}J_{0}\left(q\left(t+\frac{L}{2}\right)\sin\alpha\right)\frac{J_{1}(v)}{v}
\end{equation}

where $u=qR\cos\alpha$ and $v=q(R+t)\cos\alpha$. $\rho_{core}$,
$\rho_{shell}$, and $\rho_{med}$ are the SLDs of the core, the shell,
and the medium, respectively, and $J_{0}(x)=\frac{sin(x)}{x}$.

\subsection{Magnetic form factors $F_{M}(q)$}

The magnetic form factor $F_{M}(\vec{q})$ for a magnetic atom is
defined as:
\begin{equation}
F_{M}(\vec{q})=\frac{\gamma r_{0}}{2V}\intop\vec{\sigma}\cdot\vec{M}_{\perp}(\vec{r})e^{i\vec{q}.\vec{r}}d\vec{r}
\end{equation}
where $r_{0}=\frac{e^{2}}{m_{e}c^{2}}$ and $\gamma/2=\lyxmathsym{\textminus}1.91$
are respectively the electron radius and the Land\textasciiacute{}e
factor for neutrons. $\vec{\sigma}$ is the Pauli operator of $s=1/2$
neutrons. $F_{M}(\vec{q})$ is thus proportional to the Fourier transform
of the component of $\vec{M}$ perpendicular to the scattering vector
$\vec{q}$, that is the magnetization component parallel to the nanowire
axis in the particular case where the nanowires are oriented along
the incident beam direction $\vec{k}_{i}$ ($\vec{q}$ is essentially
normal to $\vec{k}_{i}$). In analogy with the nuclear SLD, the magnetic
SLD for an assembly of magnetic atoms can be written:

\begin{equation}
\rho_{M}=\frac{e^{2}\gamma}{2mc^{2}}\sum_{i}c_{i}M_{i}^{\perp}
\end{equation}

where $c_{i}$ is the atomic concentration of the ith species and$\frac{e^{2}\gamma}{2mc^{2}}=0.27\times10^{\text{\textminus}12}$
cm. The definition of the magnetic form factor for one atom {[}Eq.
(10){]}, can be extended to a magnetic particle of volume $V_{i}$
by introducing the magnetic contrast density between the magnetic
particles and their a priori nonmagnetic surrounding medium $\Delta\rho_{M}=\rho_{M,i}\lyxmathsym{\textminus}\rho_{M,med}$:
\begin{equation}
F_{M}(\vec{q})=\intop_{V_{i}}\Delta\rho_{M}e^{i\vec{q}.\vec{r}}d\vec{r}
\end{equation}
where the sum is over the particle\textquoteright{}s volume. Assuming
that the \textquotedblleft{}surroundings\textquotedblright{} are nonmagnetic,
$\rho_{M,med}=0$, then we have:
\begin{equation}
F_{M}(\vec{q})=\intop_{V_{i}}\rho_{M,i}e^{i\vec{q}.\vec{r}}d\vec{r}
\end{equation}
We now consider that the field is applied along the nanowire axis
and that the magnetic field is sufficiently strong so that the magnetization
inside the nanowires is identical for all nanowires and uniform inside
the volume of all nanowires ($\rho_{M,i}=\rho_{M},\forall i$). This
assumption seems perfectly valid at high fields (above 0.5 T according
to Fig. 2) but may loose its validity at much smaller fields. Then
we can take the SLD term $\rho_{M}$ out of the integral and we obtain:
\begin{equation}
F_{M}(\vec{q})=\rho_{M}\intop_{V_{i}}e^{i\vec{q}.\vec{r}}d\vec{r}=\rho_{M}F_{geo}(\vec{q})
\end{equation}
where $F_{geo}(\vec{q})$ depends solely on the particle\textquoteright{}s
shape. $F_{M}(\vec{q})$ and $F_{N}(\vec{q})$ are thus related as:
\begin{equation}
F_{M}(\vec{q})=\frac{\rho_{M}}{\Delta\rho}\frac{V_{i}}{V_{p}}F_{N}(\vec{q})=\chi(\vec{q})F_{N}(\vec{q})
\end{equation}

where $V_{i}$ and $V_{P}$ are the particle\textquoteright{}s magnetic
volume and the particle's structural volume, respectively. We make
here this distinction to emphasize the difference between magnetic
neutron scattering (sensitive to $V_{i}$) and nuclear neutron scattering
(sensitive to $V_{P}$). Obviously, the parameter $\chi(\vec{q})$
(in units of $\mu_{B}$) can be simplified when identifying the structural
particle volume and its magnetic analog ($V_{P}=V_{i}=\pi R^{2}L\approx1.32\times10^{\text{-}14}$
cm$^{3}$), and with SLD\textquoteright{}s $\rho_{Co}=2.26\times10^{10}$
cm$^{-2}$ (density 8.9 g.cm$^{-3}$), $\rho_{Al_{2}O_{3}}=4.88\times10^{10}$
cm$^{-2}$ (density 3.4 g.cm$^{-3}$), and particle density $c_{i}\approx\aleph\rho_{Co}/MCo=9.08\times10^{22}$
cm$^{-3}$ ($\aleph$ is the Avogadro number), we obtain:
\begin{equation}
\chi\approx-0.936\sum_{i}\left(\frac{M_{i}^{\perp}}{\mu_{B}}\right)
\end{equation}
Various uncertainties may come into play in this estimate: pores which
may not be totally filled, the uniformity of the pore dimensions,
the effective depth of the membranes, etc., so that an error of $10\lyxmathsym{\textendash}15
$ should be considered. If the magnetization is not uniform inside
and/or outside the nanowires, e.g., due to end domains, or stray or
demagnetizing fields {[}27,48,49{]}, then one should consider $\chi(q)$
in its generality:
\begin{equation}
F_{M}(\vec{q})=\chi(\vec{q})F_{N}(\vec{q})
\end{equation}
The function $\chi(\vec{q})$ expresses the spatial distribution of
magnetization inside the sample either through SLD variations or effective
magnetized volume, and may be extremely complex in the presence of
domains, local inhomogeneities, or magnetization gradients for instance.

\subsection{Structure factor $S(q)$}

The structure factor S(q) is a consequence of the 2D periodic arrangement
of the pores/nanowires in the alumina membrane. A regular and infinite
pattern will induce \textquotedblleft{}Bragg peaks\textquotedblright{}
for $q$ values of the associated reciprocal space. SEM images of
the porous alumina membranes show a medium- to long-range triangular
array with interpore distance $d_{P}$ around 105 nm and apparent
pore diameter $\phi_{P}$ around 25\textendash{}30 nm (see Fig. 1).

$S(q)$ can be evaluated in several ways; numerically from SEM images
from which $S(q)$ is obtained by Fourier transform. We found that
a Percus-Yevick function is in good agreement with the structure factor
as obtained from SEM images. We have adapted a Percus-Yevick model
{[}50,51{]}, originally proposed to describe colloidal particles in
liquids, to model $S(q)$ including some disorder in the area of the
pores at the membrane\textquoteright{}s surface. The Percus-Yevick
structure factor $S_{PY}(q)$ is written as:

\begin{equation}
S_{PY}(q)\sim\left(1-nC\left(q\phi_{S}\right)\right)^{-1}
\end{equation}

where $\phi_{S}$ is the diameter, $n$ is a volume density, and $C\left(q\phi_{S}\right)$
is the radial distribution function. The density number $n$ is related
to the packing density parameter $\eta$ through $n=\left(6\eta/\pi\right)\phi_{S}^{-3}$,
which is physically limited by the close-packed solid value $\eta_{max}=\pi3\sqrt{2}=0.74$.
The correlation function $C\left(q\phi_{S}\right)$ is given by {[}51{]}:
\begin{equation}
C\left(q\phi_{S}\right)=-4\pi\phi_{S}^{3}\intop_{0}^{1}x^{2}J_{0}(xq\phi_{S})(\alpha+\text{\ensuremath{\beta}}x+\text{\ensuremath{\gamma}}x^{2})dx
\end{equation}
with $J_{0}(u)=\frac{sin(u)}{u}$, $\alpha=(1+2\eta)^{2}(1\lyxmathsym{\textminus}\eta)^{-4}$,
$\beta=6\eta(1+0.5\eta)^{2}(1\lyxmathsym{\textminus}\eta)^{-4}$,
and $\lyxmathsym{\textgreek{g}}=0.5\eta(1+2\eta)^{2}(1\lyxmathsym{\textminus}\eta)^{-4}$.
Obviously, a Prevus-Yevick model is not necessarily adequate to describe
an array of disks, but this model turns out to suit the case of partially
disordered porous aluminamembranes by identifying $\phi_{S}$ with
the pore diameter $\phi_{P}$. However, the Percus-Yevick model should
be applied with some caution in the present case for two reasons:
First, the packing density parameter $\eta$ should not be interpreted
as a volume occupation, but rather as an adjustable parameter. Secondly,
the validity of the model is dependent upon the quality of the pore
ordering. In fact, the more disordered the system is, the better the
model fits. For very well ordered systems (triangular with mediumrange
order in the present case), one obtains a diffraction image and the
Percus-Yevick model is then less reliable {[}27{]}. For assemblies
of pores exhibiting some orientational disorder, the Bragg spots are
no longer observed at specific $q$ vectors, but the scattering intensity
is spread along a ring of constant $|q|$. 

Another approach that takes into account the high level of structural
order is to consider a 2D triangular lattice with nearest-neighbor
interpore spacing $d_{P}$; the structure factor $S_{triang}(q)$
is written as:
\begin{equation}
S_{triang}(q)=\frac{1}{n^{2}}\sum_{p}^{n}\sum_{q}^{n}J_{0}(qd_{pq})
\end{equation}
where $J_{0}$ is the zeroth-order Bessel function. Restricting the
sum to nearest neighbors and next-nearest-neighbors on a 2D triangular
array, we have:
\begin{equation}
S_{triang}(q)=\frac{1}{361}\left[19+168J_{0}(qd_{p})+54J_{0}(2qd_{p})+60J_{0}(\sqrt{3}qd_{p})+36J_{0}(\sqrt{7}qd_{p})+24J_{0}(3qd_{p})\right]
\end{equation}
thus matching the reflection positions expected for triangular lattice:
$q_{10.0}=q_{0}=(2/\sqrt{3})2\pi/d_{p}$, $q_{00.2}=\sqrt{3}q_{0}$,
$q_{10.1}=2q_{0}$, $q_{10.2}=\sqrt{7}q_{0}$ and $q_{11.0}=3q_{0}$.

\section{Sans experiments: results}

\begin{figure}
\includegraphics[bb=20bp 0bp 452bp 595bp,clip,width=8.5cm]{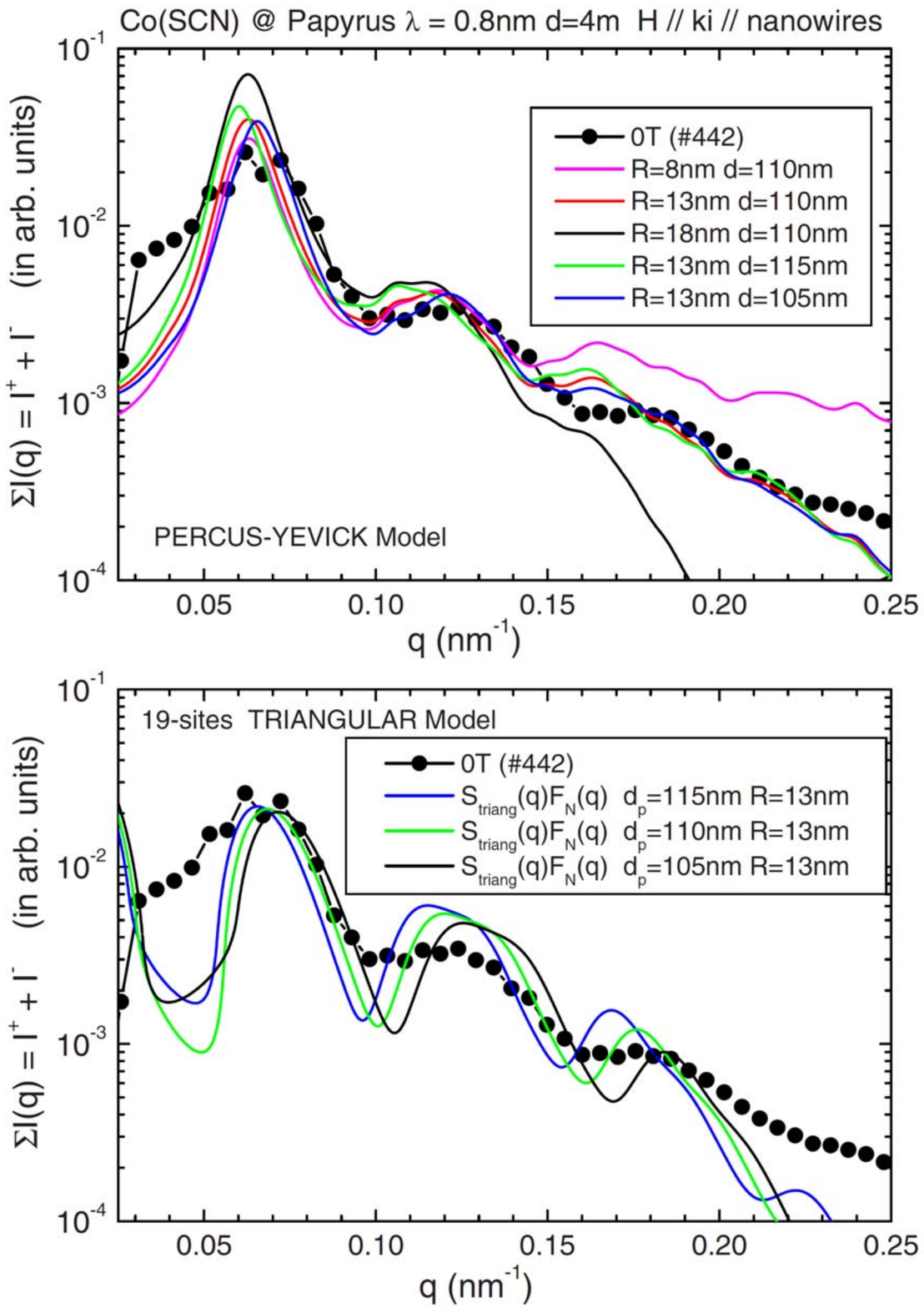}
\caption{(a) SANS data ($\lambda=$ 0.8 nm) compared with a fitted SANS model
including a Percus-Yevick structure factor SPY(q) and a Bessel-type
expression for the nuclear form factor FN(q). The best agreement is
found for $d_{P}=110$ nm and $R=13$ nm (red line). For all fits,
$\eta$ is fixed to 0.5. (b) Triangular array model {[}Eq. (21){]}
with $R=13$ nm fixed and varying interpore distance $d_{P}$. The
agreement is correct regarding peak positions but rather poor regarding
lineshapes.}
\end{figure}

The PSANS experimentswere performed on the PAPYRUS (G5.5) spectrometer
at the Laboratoire Léon Brillouin (CEA Saclay) with a neutron wavelength
of $\lambda=0.8$ nm and a sample-to-detector distance of 400 cm.
The detector is a $64\times64$ cm$^{2}$ BF$_{3}$ 2D grid with 5
mm pixel size ($128\times128$ pixels). The direct beam at the detector
(central position) is absorbed by a cadmium beam stopper. The incoming
neutrons are polarized in the up ($+$) direction (vertical direction
$y'$ as shown in Fig. 3) by a polarizing mirror to achieve a polarization
degree of $P=0.95$ of the neutrons (determined from reflectivity
curves of a reference ferromagnetic thin film). An adiabatic spin
flipper reverses the neutron polarization from up ($+$) to down ($-$),
that is from $+y$ to $-y$ . The polarization of the outgoing neutrons
hitting the detector is not analyzed. In the present PSANS experiment,
the sample (disk of $\sim8$ mm diameter)was placed perpendicular
to the incoming beam; that is, with $Z$ axis parallel to $\vec{k}_{i}$.
In such case, $\alpha=0$ and hence $\vec{q}=\vec{k}_{f}\lyxmathsym{\textminus}\vec{k}_{i}\approx\vec{q}_{\perp}$
is a very good approximation. The data are represented as a function
of $q$, which is obtained from the usual SANS expression $q=k_{i}\sin\theta$
where $k_{i}=2\lyxmathsym{\textgreek{p}}/\lambda$ and $\theta$ is
the scattering angle. Small disorientation of the nanowires can contribute
so that $\alpha$ marginally deviates from 0, meaning that the SANS
data may be dependent on the length of the nanowires. Alignment as
good as $0.5\text{\textdegree}$ is required to obtain circular \textquotedblleft{}ring-type\textquotedblright{}
SANS scattering. The horizontal magnetic field (7 T superconductor
magnet by Oxford Instruments, allowing for relatively large angle
scattering ($\pm10\text{\textdegree}$)) was set parallel to the incoming
beam and hence parallel to the nanowire axis ($\vec{H}\parallel\vec{z}$).
The transmitted beam represents only $\approx20\%$ of the incident
beam due to the very strong scattering from the sample. This confirms
that such arrays of ordered nanowires strongly interact with the incidentwave
as shown by Grigoriev et al. {[}52{]}. The intensity profile $I(q)$
was obtained after circular integration around the direct beam central
position.

Figure 4 shows SANS intensity $I(q)$ of Co nanowires embedded in
porous alumina. The first low-$q$ peaks, located at $q\approx0.065$
nm$^{-1}$ and $q\approx0.12$ nm$^{-1}$, can serve to estimate the
inter-nanowire distance $d_{P}$ and nanowire radius $R$. In a first
approximation in the triangular lattice case, $d_{P}=(2/\sqrt{3})(2\pi/0.065)\approx112$
nm, but a more careful estimate is necessary. To this end, we have
used the Percus-Yevick model {[}Eq. (18){]}. As shown in Fig. 4, the
best agreement is found for $d_{P}=110$ nm, $R=13$ nm {[}the main
effect of the radius parameter $R$ is seen by a modification of the
$I(q)$ slope at large $q${]} and a packing density parameter $\eta=0.5$
{[}53{]}. The sensitivity to the parameters indicates that errors
in the range of $10\%$ are a maximum limit. Introducing inhomogeneous
SLD across the wires did not significantly improved the results and
might lead to over-parametrization. Therefore, this model appears
sufficient to discuss the data. As for comparison between SEM and
SANS, it is worth noting that SEM introduces a Gaussian high-frequency
noise which corresponds to a spread Gaussian in the reciprocal space,
leading to a blurred zone around the holes on the SEM images and preventing
an accurate determination of the hole diameter by SEM. Therefore,
some caution should be taken when comparing SEM and SANS results.
The bottom panel of Fig. 4 shows the results of applying the 19-site
triangular 2D model {[}defined in Eq. (21){]} with a various $d_{P}$
values. The overall agreement is satisfactory concerning the peak
positions, however, the relative intensities are poorly described
by this model. The agreement in the estimate of the nanowire/pore
radius between SANS, which is a bulk technique, and a surface technique
like SEM is very good and gives some credit to the interpretation
of the SEM images presented earlier. In addition, as discussed below,
the \textquotedblleft{}magnetic\textquotedblright{} radius derived
from the present data is also in agreement with the structural radius.

\begin{figure}
\includegraphics[bb=20bp 0bp 820bp 595bp,clip,width=8.5cm]{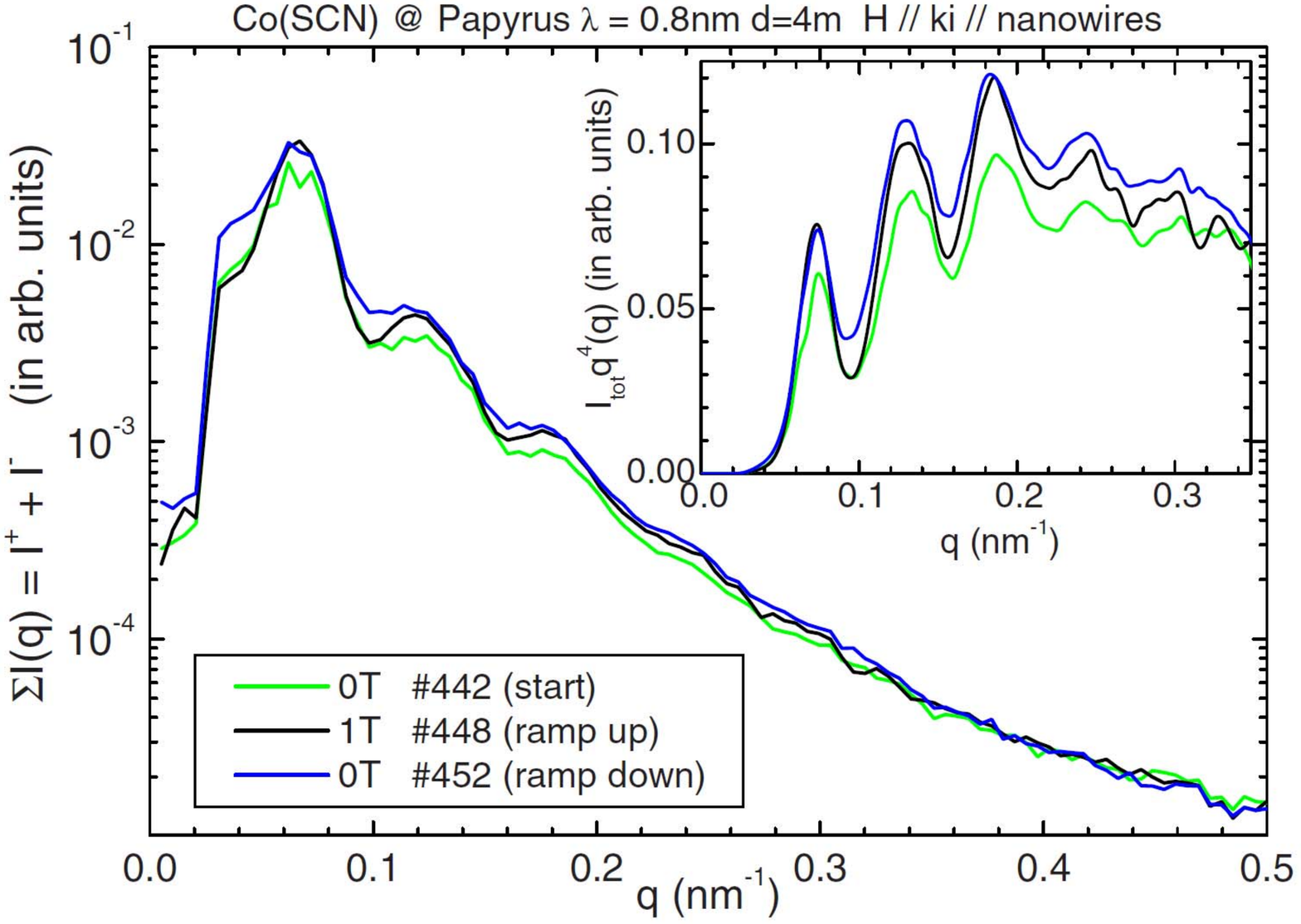}
\caption{SANS total intensities $\sum I(q)=I^{+}(q)+I^{-}(q)$ first measured
at 0 T prior to any magnetization of the sample (green curve), then
at +1 T (black curve) and finally back at 0 T (blue curve). The inset
shows the same data but represented as $\sum I(q)q^{4}$.}
\end{figure}

In order to access the magnetic behavior of the Co nanowires, we applied
magnetic fields parallel to the long axis of the nanowires and focused
on the scattering evolution. Figure 5 shows $\sum I(q)=I^{+}(q)+I^{-}(q)$
for three field values: first at 0 T prior to any magnetization of
the sample (green curve), then at 1 T (black curve), and finally at
0 T (blue curve). Before discussing the effect of the magnetic field,
let us discuss the structure factor peaks. As shown in the inset of
Fig. 5, the \textquotedblleft{}high-$q$\textquotedblright{} peaks
are much more visible in a $\sum I(q)q^{4}$ representation showing
eventual deviations from the Porod law at large $q$ originating from
interface scattering {[}54{]}. At least four peaks are clearly visible
at q = 0.073, 0.130, 0.184, and 0.244 nm$^{-1}$, and they correspond
well with the values derived from the purely triangular array structure
factor model with $d_{P}=110$ nm ($q=0.066,0.114,0.132,0.174,0.198,$
and 0.228 nm$^{-1}$). The observed peaks at $q=0.130$ and 0.184
nm$^{-1}$ are the result of two unresolved peaks at $0.11\lyxmathsym{\textendash}0.13$
and $0.17\lyxmathsym{\textendash}0.2$ nm$^{-1}$. For $d_{P}=1\text{0}5$
nm and $d_{P}=115$ nm, the agreement is significantly worse.

\begin{figure}
\includegraphics[bb=20bp 160bp 820bp 595bp,clip,width=8.5cm]{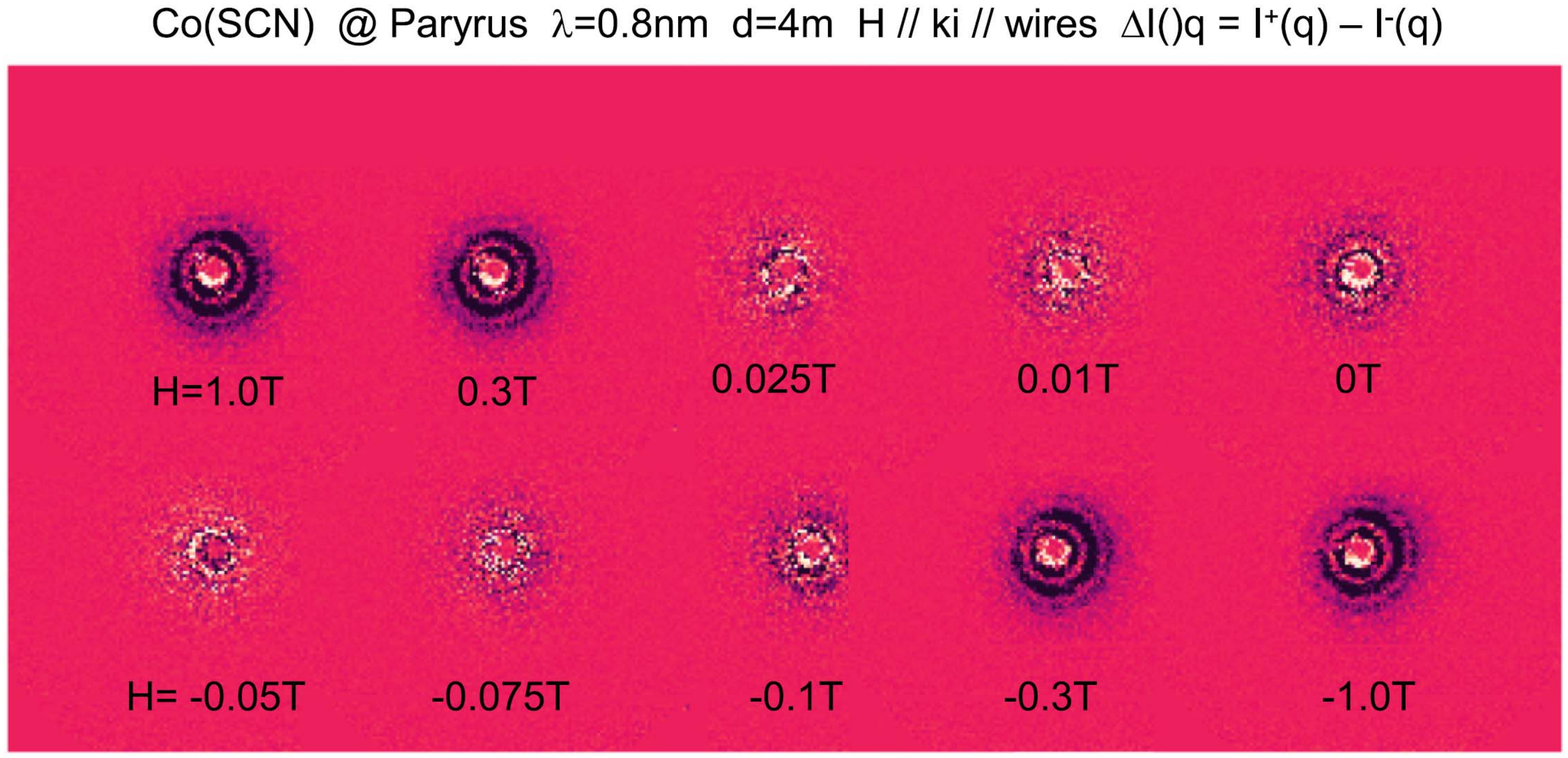}
\caption{2D detector maps representing $\Delta I(q)=I^{+}(q)+I^{-}(q)\propto F_{M}(q)$
at 200 K and for several magnetic fields $H$ parallel to the long
axis of the cobalt nanowires and to the incident beam direction.}
\end{figure}

Regarding magnetic field effects, several comments can be made. First
there is a significant increase of scattering between 0 and 1 T, characterized
by an enhancement of the structure factor peaks and some additional
intensities in between the $S(q)$ peaks from 1 T back down to 0 T,
which show the importance of magnetic history in this system. The
evolution of the magnetic scattering $\Delta I(q)=I^{+}(q)+I^{-}(q)\propto F_{M}(q)$
as a function of the applied magnetic field is first shown in Fig.
6 as 2D maps of the detector. The observed concentric rings are clearly
field dependent, with low magnitude (positive or negative) at low
fields and maximized magnitude at large magnetic fields ($\pm1$ T).
At intermediate fields, the pattern is more complex and the several
visible concentric rings are better represented after a circular integration
of the 2D plots. This is represented in Fig. 7 where the color code
groups data sets of similar q dependence: high-field ($\mu_{0}|H|>0.3$
T) data drawn in black, lowfield data are in green, and blue/red show
intermediate magnetic fields. By introducing a scaling factor $K(H)$,
$\Delta I(H)\sim K(H)\Delta I(1T)$, where the data at 1 T serve as
reference data for \textquotedblleft{}fully magnetized\textquotedblright{}
nanowires along their long axis, one can derive from these curves
a hysteretic behavior of $\Delta I(q,H)$ by plotting the scalar $K(H)$
(see inset in Fig. 7) at some chosen $q$ value {[}herewe choose $q=0.073$
nm$^{-1}$ which is the position for the largest $\Delta I(q)$ value
at 1 T{]}. While green data are characterized by very low $\Delta I(q)$
values, red and blue $\Delta I(q)$ values are relatively large and,
most noticeably, with $K(H)$ values of opposite sign. 

After noticing that the structure factor probes only nuclear densities
and, therefore, remains unchanged with magnetic field variations,
one can infer from Eq. (3) that the observed field-dependence of $\Delta I(q)$
is due to the magnetic form factor $F_{M}(q)$. The striking feature
revealed in the inset of Fig. 7 is the sign inversion of $K(H)$ occurring
at $\pm50$ mT, followed by another more modest inversion below $\pm20$
mT, which is close to the coercive field value (38 mT, see Fig. 2).
It shows that the magnetization component $M_{\perp\vec{q}}$ , through
the present evolution of $F_{M}(\vec{q})$, presents inversion features
at low fields. To go further in the analysis, it is necessary to eliminate
the structure factor $S(q)$ contribution by considering the evolution
of $\chi(\vec{q})=F_{M}(\vec{q})/F_{N}(\vec{q})$ {[}Eq. (17){]}.
From $\chi(\vec{q})$, one can extract directly $F_{M}(\vec{q})$
since we know the nuclear form factor $F_{N}(\vec{q})$ from Eq. (7)
(with $q_{\parallel}L\ll1$):
\begin{equation}
F_{M}(\vec{q})=\chi(\vec{q})F_{N}(\vec{q})=-0.936\sum_{i}\left(\frac{M_{i}^{\perp}}{\mu_{B}}\right)\left(\Delta\rho\right)V_{P}\frac{2J_{1}(q_{\perp}R)}{q_{\perp}R}
\end{equation}

\begin{figure}
\includegraphics[bb=20bp 10bp 820bp 595bp,clip,width=8.5cm]{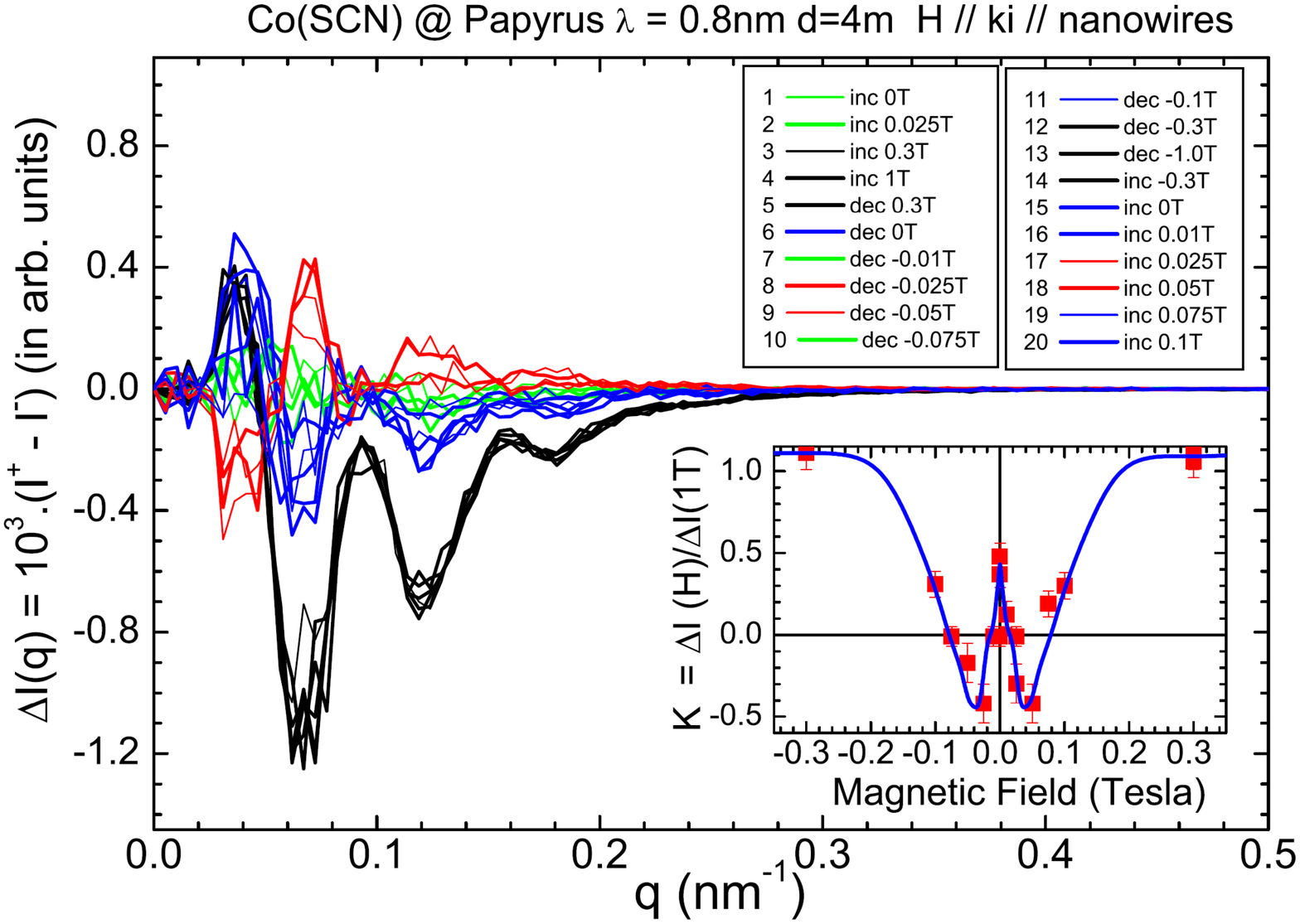}
\caption{Circular integration of PSANS intensities $\Delta I(q)=I^{+}(q)+I^{-}(q)\propto F_{M}(q)$.
The color code groups data sets with the same \textquotedblleft{}shape\textquotedblright{}
or $q$ dependence. High-field ($\mu_{0}|H|>0.3$ T) data are in black;
low-field data are in green and blue/red show intermediate magnetic
fields. As in Fig. 5, the peaks are located at the same q values.
The $q$-oscillatory behavior can be qualitatively expressed as $\Delta I(H)\sim K(H)\Delta I(1T)$
where 1 T data serve a benchmark. While green data are characterized
by very low $\Delta I(q)$ values, red and blue $\Delta I(q)$ values
are relatively large with reversed sign of $K$. The inset shows the
proportionality term $K(H)$ as a function of magnetic field (hysteresis
loop between -1 T and +1 T) obtained from the value of $\Delta I(q)$
at $q$ = 0.073 nm$^{-1}$ position for each magnetic field value.
The solid line is a guide to the eyes.}
\end{figure}

Knowing $\chi(\vec{q})$ from the PSANS experiments and $F_{N}(\vec{q})$
(best fits to the unpolarized data and SEM images with $R=13$ nm
as the main parameter), we can plot the magnetic form factor $F_{M}(\vec{q})$
(see Fig. 8) for three magnetic field values (in chronological order
1 T, 25 mT, and 0 T) corresponding to the three regimes identified
in Fig. 7.We then make the assumption that $F_{M}(\vec{q})$ can be
expressed as the product of an amplitude scaling parameter ($K'$)
and a geometrical magnetic form factor which represents the magnetic
\textquotedblleft{}landscape\textquotedblright{} of the sample: $F_{M}(\vec{q})=K'F_{M}^{geo}(\vec{q})$.
We find that the geometrical magnetic form factor $F_{M}^{geo}(\vec{q})$
is best modeled using a \textquotedblleft{}core-shell cylinder\textquotedblright{}
type geometrical magnetic form factor {[}see Eq. (9){]}:
\begin{equation}
F_{M}^{geo}(\vec{q})=I_{1}\frac{J_{1}(qR_{1})}{qR_{1}}+I_{2}\frac{J_{1}(qR_{2})}{qR_{2}}
\end{equation}

\begin{figure}
\includegraphics[bb=20bp 10bp 820bp 595bp,clip,width=8.5cm]{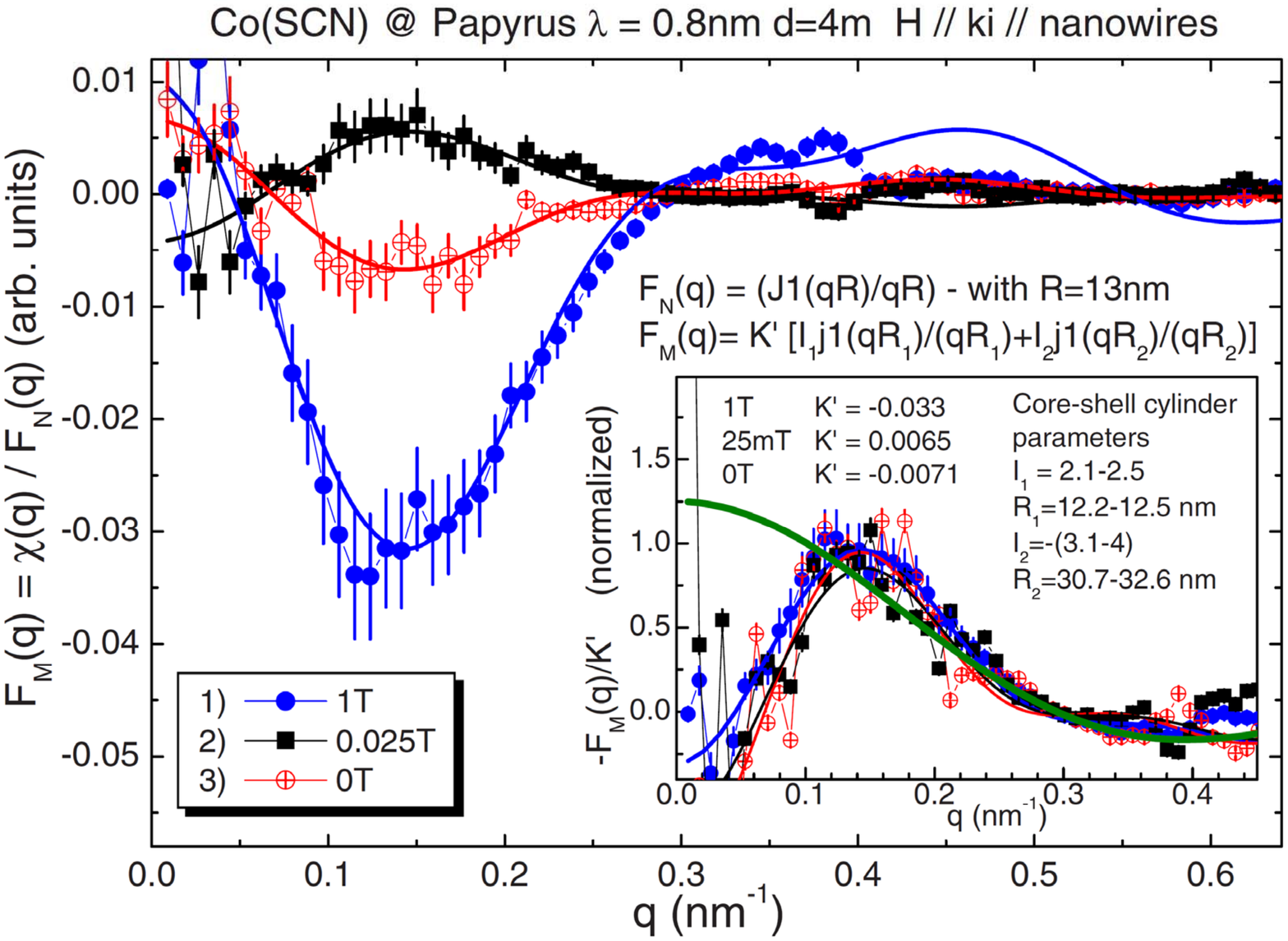}
\caption{Magnetic form factor $F_{M}(\vec{q})$ of ferromagnetic Co nanowires
for different longitudinal magnetic field fields: In chronological
order (1) 1 T, (2) 25 mT, and (3) 0 T. Solid lines are best fits using
a core-shell cylinder model with the parameters $I_{1}=2.3\pm0.2$,
$R_{1}=12.35\pm0.15$ nm, $I_{2}=3.6\pm0.5$, and $R_{2}=31.6\pm0.9$.
Inset: $F_{M}(\vec{q})$ normalized to unity (divided by $K'$). The
scaling parameter $K'$ for each magnetic field is indicated. The
green continuous line in the inset panel represents the expected magnetic
form factor in the absence of a dipolar shell (i.e., with $I_{2}=0$).}
\end{figure}

where $R_{1}$ is the core radius and $R_{2}=R_{1}+t$ is the shell
radius. The results are shown in Fig. 8. A very good agreement is
found in the three identified regimes with the following parameters:
with $I_{1}=\Delta\rho_{1}V_{1}=2.3\pm0.2$, $R_{1}=12.35\pm0.15$
nm ($V_{1}\approx500$), $I_{2}=\Delta\rho_{2}V_{2}=3.6\pm0.5$, $R_{2}=31.6\pm0.9$
nm ($V_{2}\approx2500$), and $K'$ values as shown in the inset of
Fig. 8. Setting arbitrarily $I_{2}=0$ or $I_{2}>0$ leads to a monotonic
decrease of $F_{M}^{geo}(\vec{q})$, in total disagreement with the
experimental observation. As an example, the bold green line in Fig.
8 shows the magnetic form factor for $I_{2}=0$. Qualitatively, the
position of the main oscillation in $F_{M}^{geo}(\vec{q})$ is set
by $\approx2\pi/R_{2}$. The $R_{1}$ and $R_{2}$ values deserve
comment. $R_{1}$ is extremely close to the structural nanowire/pore
radius (13 nm), indicating that the \textquotedblleft{}core\textquotedblright{}
magnetization inside the wire extends across all the nanowire volume
and that the magnetization is essentially uniform, even for low magnetization
values. The \textquotedblleft{}shell\textquotedblright{} radius, much
larger than the structural nanowire radius, $R_{2}\approx31$ nm,
reflects the fact that dipolar fields (opposed to the core magnetization)
extend in between nanowires. The dipolar field profile $\mu_{0}\vec{H}_{dip}(\vec{q})$
may differ strongly depending on the length of the nanowires and on
the internanowire distance, the type of packing, the internal magnetization
value, etc. {[}20{]}. The scaling parameter $K'$ exhibits a surprising
field dependence with inversion features at low fields (below 0.1
T) that, seemingly, is not related to the internal magnetization distribution
inside/outside the nanowires.

\begin{figure}
\includegraphics[bb=20bp 10bp 820bp 595bp,clip,width=8.5cm]{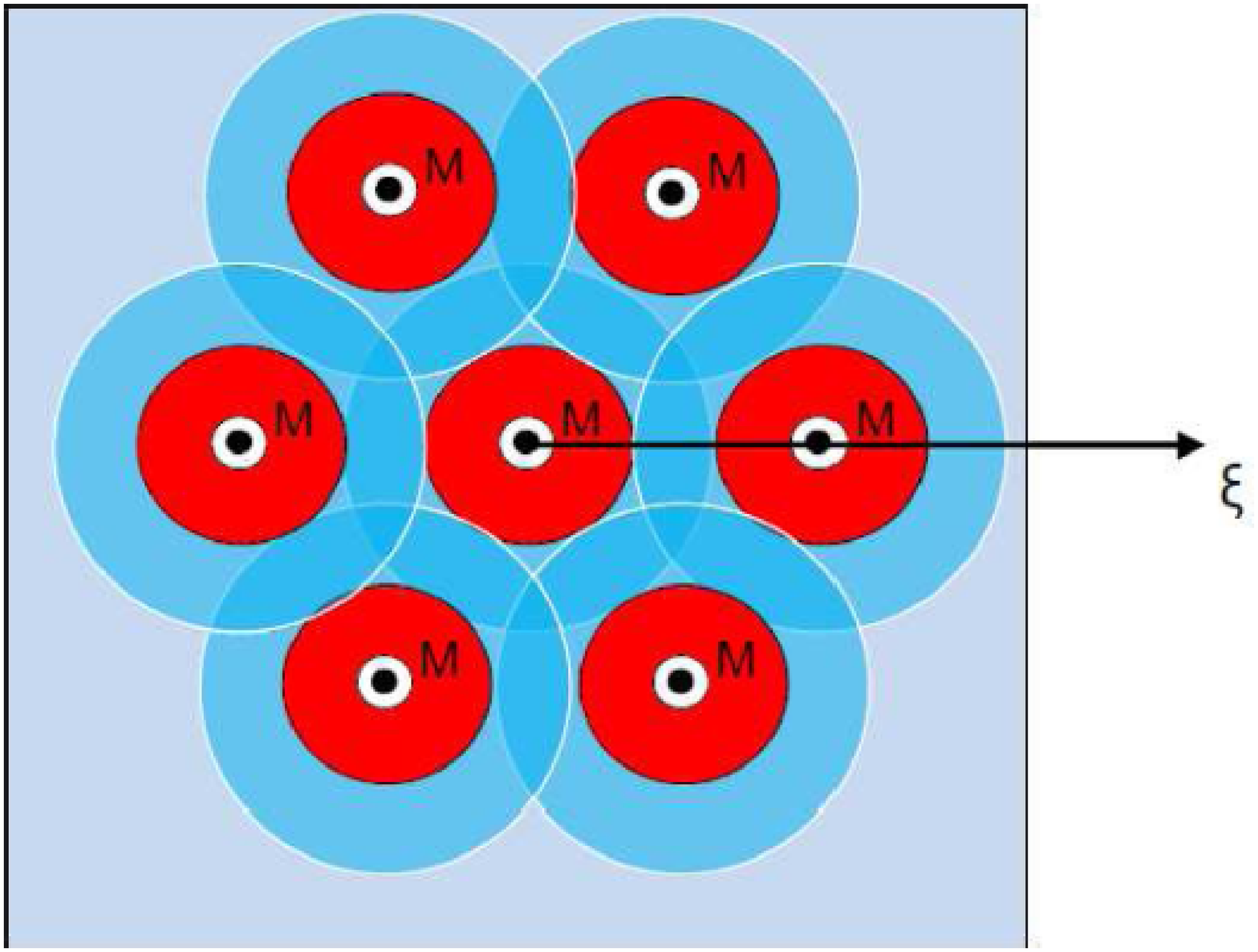}
\caption{Triangular array of nanowires with homogeneous magnetization inside
the core (in red) and dipolar field intensity (in blue). The dipolar
field creates a neutron scattering contrast with the non-magnetic
matrix. The axis $\xi$ will meet different dipolar field profile
depending on its relative orientation. The data obtained will then
correspond to a profile average.}
\end{figure}

\begin{figure}
\includegraphics[bb=20bp 200bp 820bp 595bp,clip,width=8.5cm]{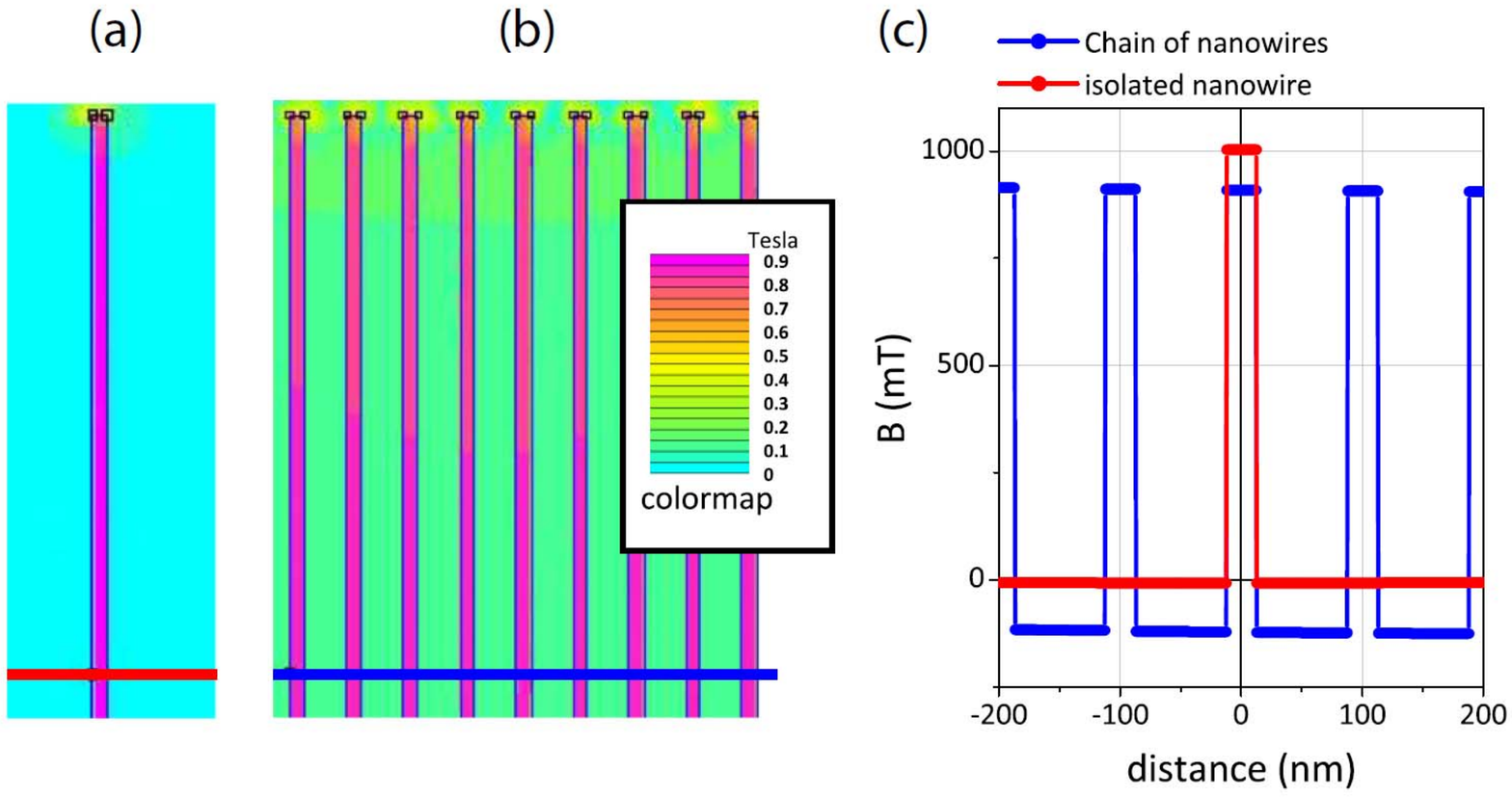}
\caption{(a) Dipolar fields generated by an individual nanowire with a diameter
of 25 nm and a length of 5 $\mu$m. (b) Dipolar fields generated for
an assembly of nanowires and internanowire distance, center to center,
of 100 nm. (c) Profile of the induction for both situations, indicating
that the dipolar fields are sizable in magnitude (with reversed sign)
relative to the core induction. In the 2D triangular situation {[}24{]}
the effect is further enhanced. The simulation has been performed
with the FEMM software.}
\end{figure}

From the core-shell model, one can estimate the magnetic moment induced
by the nanowires in the matrix volume around the nanowires:

\begin{equation}
\gamma_{1,2}=\frac{\rho_{M,1}-\rho_{M,2}}{\rho_{M,2}}=\frac{I_{1}}{I_{2}}\times\frac{V_{2}}{V_{1}}
\end{equation}

where the subscripts 1 and 2 relate to the nanowire core and to the
outer dipolar field volume, respectively. With $\frac{I_{1}}{I_{2}}\approx-0.638$
and $\frac{V_{2}}{V_{1}}\sim\left(\frac{R_{2}}{R_{1}}\right)^{2}\sim6.54$,
we obtain $\rho_{M,1}/\rho_{M,2}=1+\gamma_{1,2}\approx-3.17$. The
quantity $\rho_{M,1}$ is related to the magnetization component $M_{i}^{\perp}$
of the Co atoms, which is, at full saturation, $M_{Co,sat}\approx1.7\lyxmathsym{\textendash}1.75$
$\mu_{B}$/(Co atom) $\approx$ 1400 kA/m, equivalent to a magnetic
field $\mu_{0}H{}_{core}^{z}\approx1.7$ T. From the analysis of the
experimental data, we would obtain an opposing magnetic field in the
shell region around the nanowires of $\mu_{0}H{}_{shell}^{z}=-\rho_{M,2}/\rho_{M,1}\times\mu_{0}H{}_{core}^{z}$
$\lyxmathsym{\textminus}0.315\times1.7\approx\lyxmathsym{\textminus}-0.53$
T, as depicted in Fig. 9. In some instances, the demagnetization field
can be phenomenologically related to the \textquotedblleft{}porosity\textquotedblright{}
$P$ of the array {[}55,56{]}: $H_{d}\approx\lyxmathsym{\textminus}M_{Co,sat}P$.
This is equivalent to saying that the demagnetizing field of a regular
array of ferromagnetic nanowires is the demagnetizing field of a uniform
ferromagnetic film modulated by a porosity factor (surface ratio).
With $P=\pi\phi_{P}^{2}/(2\sqrt{3}d_{P}^{2})\approx0.05$, we have
$\mu_{0}H{}_{d}\approx-85$ mT, which is one order of magnitude lower
than that derived from PSANS. To back up our findings, we have performed
numerical simulations, using the FEMM (Finite Element Method Magnetics)
software {[}57{]}, on nanowires ($2R=25$ nm, $L=5$ $\mu$m). The
FEMM results are shown in Fig. 10: Panel (a) shows a color map of
the dipolar field generated by one individual nanowire and a transverse
cut of the induction related to the nanowire integrated along $z$,
arising from bothmagnetization and dipolar fields. Panel (b) shows
the same features for a chain of nanowires separated by $4R$. These
results clearly indicate that dipolar fields cannot be neglected in
the calculation of the induction for an assembly of nanowires. The
intensity and local variations of dipolar fields in between nanowires
should be taken into account when it comes to evaluating the magnetic
SANS signal. These experimental results confirm the conclusion of
a recent study on 2D arrays of ordered Ni nanowires probed by SANS
{[}27{]} and performed on similar systems but with a magnetic field
transverse to the nanowires. They also exhibited results which cannot
be explained without considering complex dipolar fields, not only
at the end tips of the nanowires but also in between the nanowires
($H_{d}/H_{core}\approx-0.31$).

\section{Conclusion}

We have performed polarized small-angle neutron scattering (PSANS)
on ordered arrays of Co magnetic nanowires. PSANS is a powerful but
emerging technique which has only recently been used to investigate
the magnetic configuration of nanoparticles. For instance, PSANS revealed
chemically uniform, but magnetically distinct, core and canted shell
in 9 nm magnetite particles {[}58{]}. Here, we show that PSANS is
a tool to characterize, both structurally and magnetically, anisotropicmagnetic
nano-objects.With unpolarized neutrons, it is possible to disentangle
the structure factor of the array and the nuclear form factor of a
single wire. In previous studies, focused on Co and Ni nanowires ordered
in Al$_{2}$O$_{3}$ membranes but with the external magnetic field
applied perpendicular to the wire long axis {[}25,27,48{]}, the variation
of the SANS intensity depending on the applied magnetic field reveals
that stray fields have to be taken into account in the magnetic formfactor
derivation. In the polarized-neutrons case, we show that it is possible
to derive directly from experiment the magnetic form factor, and therefore
the magnetization distribution in the sample. We argue that it necessitates
the introduction of significant magnetostatic fields in between nanowires,
whose effects are modeled using a straightforward core-shell model
comprising (1) a core magnetic radius of induction close to the Co
value and equal to the structural radius, and (2) a \textquotedblleft{}dipolar
shell\textquotedblright{} induction of constant but opposite amplitude
surrounding the core induction. The evolution of these fields as a
function of external magnetic field is also reported. Subtle inversion
effects at very low fields due to the interplay of internal and external
fields have been evidenced. We show that a magnetization \textquotedblleft{}mapping\textquotedblright{}
in such types of nano-objects is indeed possible using PSANS techniques
and may easily be extended to further deposited nano-objects (dots,
wires, etc.). Such type of studies indicates that PSANS is a promising
technique able to provide information about complex magnetization
in nano-objects.


\begin{thebibliography}{References}
\bibitem{key-49}This is in contrast to Co nanowires grown under pH
= 3.4 (resp. pH = 5.75) which crystallize in the hcp phase with the
c axis perpendicular (resp. parallel) to the nanowire axis (z axis).
\bibitem{key-56}$\eta=0.5$works quite well, but relatively good
agreement can be found with either \textgreek{h} = 0.4 or \textgreek{h}
= 0.6.
\end{thebibliography}
\end{document}